\documentclass[article]{jdssv}
\usepackage{amsmath,amssymb}
\usepackage{graphicx}%
\usepackage{bm}%
\usepackage{hyperref}%
\usepackage{siunitx}%
\usepackage{physics}
\usepackage{xfrac}
\usepackage{csquotes}
\usepackage[caption=false,farskip=0pt,nearskip=0pt,captionskip=0pt]{subfig}%

\newcommand{\vect}{\bm}

\newcommand{\expect}{\mathbb{E}}

\DeclareMathOperator{\variance}{Var}

\author{Lukas Koch\\Johannes Gutenberg University Mainz}
\Plainauthor{Lukas Koch}

\title{Plotting Correlated Data}
\Plaintitle{Plotting Correlated Data}
\Shorttitle{Plotting Correlated Data}
\Volume{VI}
\Year{2026}
\Month{May}
\Issue{II}
\Submitdate{2026-02-02}
\Acceptdate{2026-04-19}
\DOI{10.52933/jdssv.v6i2.170}

\Abstract{
    A very common task in data visualisation is to plot many data points with some measured $y$-values as a function of fixed $x$-values.
    Uncertainties on the $y$-values are typically presented as vertical error bars that represent either a Frequentist confidence interval or Bayesian credible interval for each data point.
    Most of the time, these error bars represent a 68\% confidence/credibility level,
    which leads to the intuition that a model fits the data reasonably well
    if its prediction lies within the error bars of roughly two thirds of the data points.
    Unfortunately, this and other intuitions no longer work when the uncertainties of the data points are correlated.
    In those cases, the error bars only show the square root of the diagonal elements of the covariance matrix.
    If the covariance has non-negligible off-diagonal elements,
    we simply do not have enough information in the plot to judge whether a drawn model line agrees well with the data or not.
    In this paper, we will demonstrate this problem and discuss ways to add more information to the plots to make it easier to judge the agreement between the data and some model prediction in the plot,
    as well as glean some insight where the model might be deficient.
    This is done by adding information about the correlation between neighbouring bins, and showing the contribution of the first principal component of the uncertainties.
}

\Plainkeywords{data visualisation, correlations}

\Address{
  Lukas Koch\\
  Institute of Physics -- ETAP\\
  Johannes Gutenberg University Mainz\\
  Staudingerweg 7, 55128 Mainz \\
  E-mail: \email{lukas.koch@uni-mainz.de}
}

\begin{document}

\section{Introduction}

The visualisation of data with uncertainties is non-trivial and is an active field of research (see e.g., \cite{Kamal2021} for an overview).
The choice of a \enquote{best} visualisation technique can depend on the type of data, as well as the nature of the uncertainties, and the intended audience of the visualisation.
In this work we will focus on a common task in most quantitative sciences:
the plotting of many data points with some measured $y$-values as a function of fixed $x$-values, intended for consumption by other researchers.

In these cases, uncertainties on the $y$-values are typically presented as vertical error bars that represent either a Frequentist confidence interval or Bayesian credible interval for each data point.
Most of the time, these error bars represent a 68\% confidence/credibility level,
which leads to the intuition that a model fits the data reasonably well
if its prediction lies within the error bars of roughly two thirds of the data points.
Unfortunately, this and other intuitions no longer work when the uncertainties of the data points are correlated.
In those cases, the error bars only show the square root of the diagonal elements of the covariance matrix.
If the covariance has non-negligible off-diagonal elements,
we simply do not have enough information in the plot to judge whether a drawn model line agrees well with the data or not.

An example of this is shown in \autoref{fig:diag}.
The models M1 and M2 both stay within the error bars of the data points, so they both seem compatible with the data.
M2 is closer to the central value of the data points for all three points, so it looks like it should have even better compatibility with the data than M1.
But when calculating the actual squared Mahalanobis distances (M-distance, commonly called \enquote{the chi-squared}) between the models and the data using the data covariance, it is clear that model M2 is a much worse description of the data than M1.
This is of course due to the correlations in the data uncertainties, which are not shown in the plot.

It is clear that just plotting the usual error bars is not sufficient to judge whether a model prediction matches the data or not.
In such cases, at the very least, the squared M-distance of the model and the \enquote{degrees of freedom} need to be included in the plots to convey the actual goodness of fit.
Alternatively, the corresponding $p$-value can be shown directly.
Only showing the \enquote{reduced chi-squared}, where the squared M-distance is divided by the degrees of freedom, is not sufficient since it does not uniquely correspond to a $p$-value.

\newcommand{\pw}{0.48\textwidth}
\begin{figure}
    \centering
    \includegraphics[width=\pw]{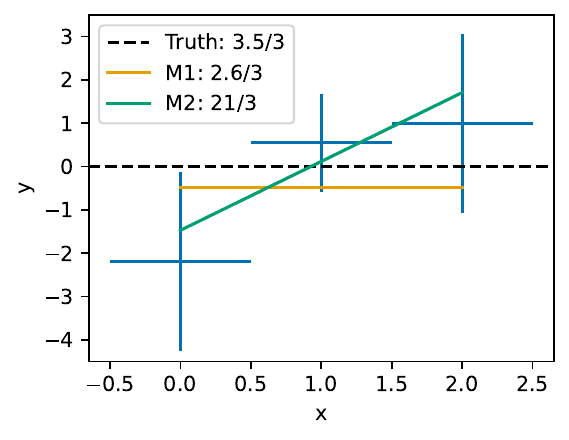}
    \caption{\label{fig:diag} Some correlated data as it is usually presented. The vertical error bars represent the square root of the diagonals of the covariance matrix. Without any additional context, the model prediction M2 looks like it is describing the data much better than the model prediction M1. Only a look at the actual squared M-distances and degrees of freedom in the legend reveals that M2 describes the data much worse.}
\end{figure}

\section{Plotting the Correlation Matrix}

\begin{figure*}
    \centering
    \subfloat[coolwarm]{\includegraphics[width=0.49\linewidth]{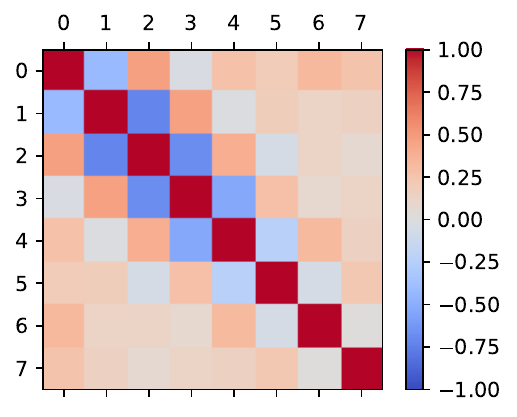}\label{fig:coolwarm}}
    \subfloat[coolwarm converted to grayscale]{\includegraphics[width=0.49\linewidth]{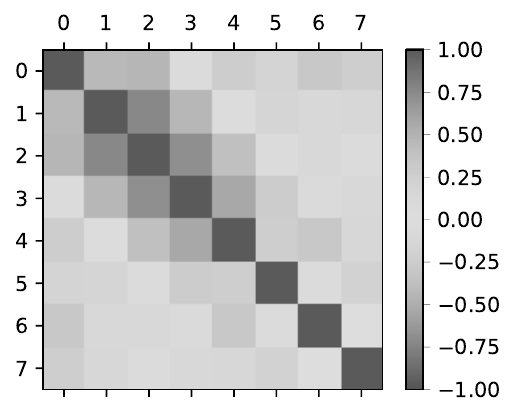}\label{fig:coolwarm-gray}}\\[1em]
    \subfloat[cividis]{\includegraphics[width=0.49\linewidth]{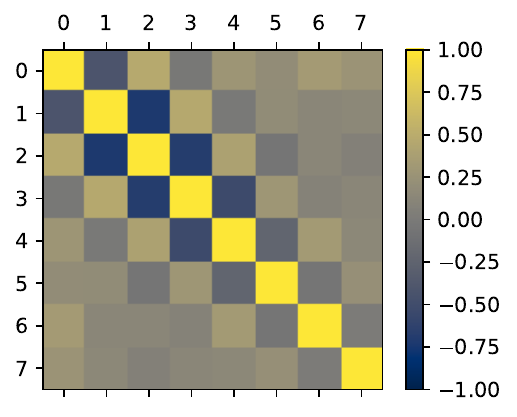}\label{fig:cividis}}
    \subfloat[gray]{\includegraphics[width=0.49\linewidth]{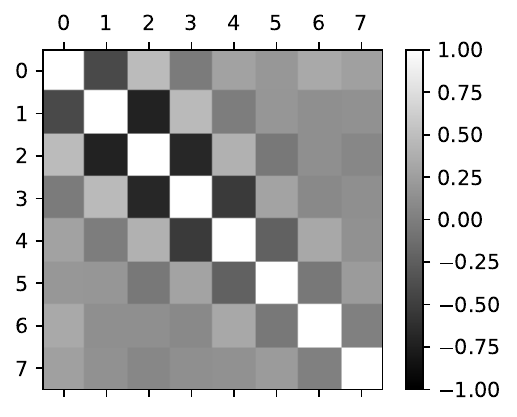}\label{fig:gray}}
    \caption{\label{fig:cor}
        Examples of colour maps used to plot correlation matrices and their names in the \pkg{Matplotlib} library \citep{MDT2025}.
        Divergent maps like \protect\subref{fig:coolwarm} show a clear distinction between positive and negative values, but they do not work without the colour information \protect\subref{fig:coolwarm-gray}.
        Perceptually uniform sequential colour maps like \protect\subref{fig:cividis} and \protect\subref{fig:gray} work well even in monochrome, but they make it harder to distinguish small positive from small negative values.
        The correlation matrix is taken from the $\delta p_T$ result in \cite{Abe2018,Abe2018e}.%
    }
\end{figure*}

The first thing one can and should do to show the correlation information that is missing in the \enquote{diagonals only} plot, is to simply plot the correlation matrix.
This is very often done as a 2D histogram with a divergent colour map.
In such plots, matrix elements with a value of 0 are represented with a white or light gray colour and positive and negative values are shown as two different colour hues that get darker with increasing absolute value (see e.g., \autoref{fig:coolwarm}).
This makes it relatively easy to distinguish positive from negative correlations, as long as one can use the colour hue information in the plot.

But because both positive and negative numbers get darker with increasing absolute value, these colour scales are not ideal from an accessibility point of view.
If the colour hue information cannot be used -- either because the plots were printed out on a b/w printer, or because the reader is colour blind -- these plots become unusable, as shown in \autoref{fig:coolwarm-gray}.

Sometimes perceptually uniform sequential colour maps are used to avoid this issue.
These colour maps use colour hue only as an additional signal (e.g., \autoref{fig:cividis}) or not at all (e.g., \autoref{fig:gray}), and all information is also available in the lightness of the colour, and positive and negative values are on opposite ends of this spectrum.
This makes them work also in instances where only the lightness information is available, but it becomes difficult to tell apart small positive from small negative values, which can be of interest for correlation matrices.

\begin{figure*}
    \centering
    \subfloat[\label{fig:hinton-civ}cividis]{\includegraphics[width=0.49\linewidth]{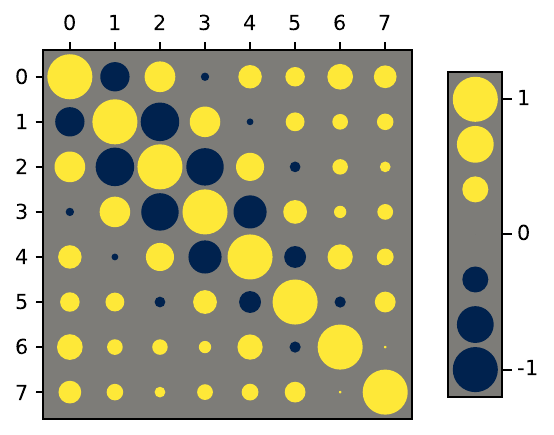}}
    \subfloat[\label{fig:hinton-gray}gray]{\includegraphics[width=0.49\linewidth]{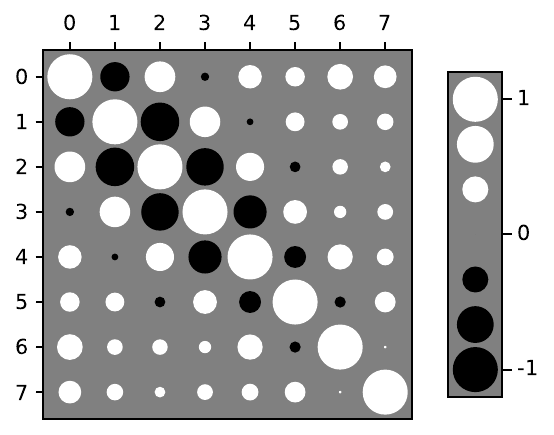}}
    \caption{\label{fig:hinton}%
        Examples of Hinton plots and the names of the used colour maps in the \pkg{Matplotlib} library \citep{MDT2025}.
        In these, the absolute value of the matrix elements is reflected as the area of the circles,
        while the colour of the circles represents the sign of the value.
        Positive values are easily distinguishable from negative values even for small absolute values and a complete lack of colour information.
        The correlation matrix is taken from the $\delta p_T$ result in \cite{Abe2018,Abe2018e}.%
    }
\end{figure*}

Hinton diagrams -- originally introduced to show weights in a neural network \citep{Hinton1991} -- can overcome this issue.
In these diagrams, the absolute value of the matrix elements is not represented by the hue or lightness of the colour, but by the area of a symbol drawn at each position corresponding to a matrix element.
The background colour is taken from the middle of the colour scale, and the colour of the symbols from both ends.
One indicates positive values and one negative (see \autoref{fig:hinton}).
Small bright dots are very easily distinguished from small dark dots.
When using circles as symbols, Hinton diagrams work well with surprisingly big matrices.
As the circles get smaller and smaller, the overall impression of the diagram becomes like a regular 2D histogram where the colour determines the element value, since it effectively turns into a halftone image \citep{Campbell2000}.

\begin{figure*}
    \centering
    \raisebox{-0.5\height}{\includegraphics[width=\pw]{simple-diag.pdf}}
    \raisebox{-0.5\height}{\includegraphics[width=0.39\linewidth]{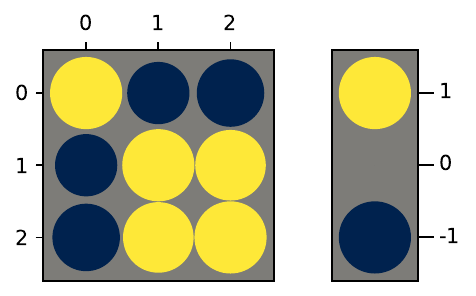}}
    \caption{\label{fig:diag-cor} The example data with error bars representing the square root of the diagonal elements of the covariance matrix together with its correlation matrix. The strong correlations visible in the matrix explain the very bad performance of the model M2. This combination of plots contains all the information available about the uncertainties, but it is inconvenient that it is spread over two separate plots. Especially with larger number of data points, it can get difficult to interpret the correlation matrix in terms of what it means for the allowed variations in the data plot.}
\end{figure*}

\autoref{fig:diag-cor} shows the example data next to its correlation matrix as a Hinton diagram.
With this additional information, one can now see that the second and third bin are strongly positively correlated, while model M2 predicts values on opposite sides of the two data points.

The presence of the correlation matrix makes it possible to understand why a model fits better or worse to the data than expected, but spreading the information over two separate plots is not ideal.
Especially with higher number of data points, it can become quite bothersome to try to assign the right matrix element to any given combination of data points, while scanning the actual data plot for differences between the data and the models.
It would be more convenient to embed some information about the correlations in the main plot itself.

\section{Correlations between Neighbouring Data Points}

For $N$ data points, there are $N(N-1)/2$ correlation coefficients, which are free to vary independently with the boundary condition that the correlation matrix remains positive (semi-)definite.
It might not be possible to display all of this information efficiently in a plot like \autoref{fig:diag}.
There is, however, space in between the data points that can be used to at least show the correlation between \emph{neighbouring} data points.

\begin{figure*}
    \centering
    \subfloat[Correlation lines]{\includegraphics[width=\pw]{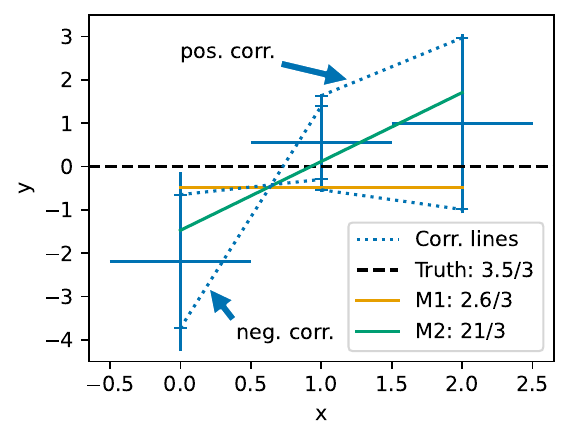}\label{fig:corlines}}
    \quad
    \subfloat[Random throws]{\includegraphics[width=\pw]{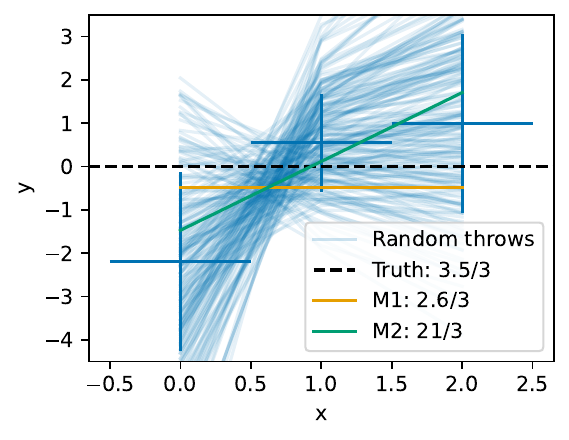}\label{fig:parallel}}
    \caption{\protect\subref{fig:corlines} The example data with correlation lines indicating the correlation between neighbouring data points.
    The position where the correlation lines connect to the error bars relative to the length of the error bars corresponds to the absolute value of the correlation coefficient between the two bins. Positive correlation is indicated by lines both connecting on the same side of the data points, anticorrelation by crossing lines which connect to opposing sides.
    If there is no correlation, there will be just a single line connecting the data points directly.
    Compared to the plot of random samples from the covariance as shown in \protect\subref{fig:parallel}, the correlation lines introduce much less clutter.
    }
\end{figure*}

One possible way of doing that is shown in \autoref{fig:corlines}.
It introduces \enquote{correlation lines}, two of which each connect neighbouring data points.
They do not connect to the data point itself, but instead attach to the vertical error bars at a relative height that corresponds to the correlation coefficient.
If the correlation is positive, the lines connect to the same sides of the two data points.
If the correlation is negative, the lines connect to the opposite sides, meaning they will cross in between the data points.
If there is no correlation, the two lines coincide and simply connect the data points directly.

These lines are similar in concept to parallel coordinates plots (see e.g., \cite{Heinrich2012}).
But where the latter can be used to show arbitrary relations between dimensions of \emph{samples} from a random distribution, the correlation lines show a single summary statistic of the relation between neighbouring dimensions in the \emph{whole distribution}.
For multivariate normal distributed data, the correlation lines convey all information with much less visual clutter than classical parallel coordinate plots, as shown in \autoref{fig:parallel}.

\begin{figure*}
    \centering
    \includegraphics[width=\pw]{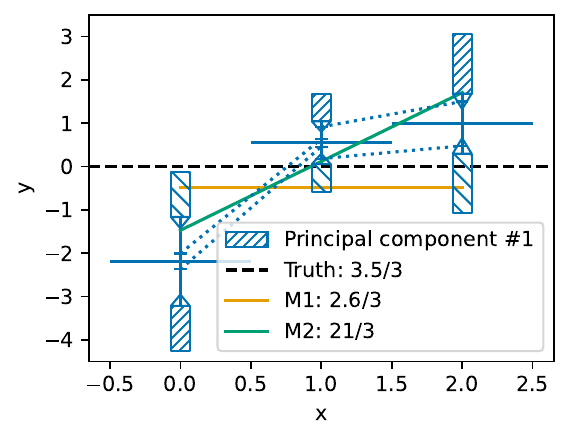}
    \caption{\label{fig:pcplot}The example data as a principal component plot.
    The hatched areas indicate that the principal component of the covariance strongly correlates the last two bins (as the hatch stiles match on the same sides of the data points) and anticorrelates them with the first bin (as the hatch styles match on opposing sides of the data points).
    When this component is reduced, the remaining covariance has positive correlations between all neighbours as indicated by the correlation lines.
    The prediction of model M2  does not align with any of those correlations.
    See \autoref{fig:full} for a detailed explanation of the different features in the plot.
    }
\end{figure*}

Aside from giving an impression of the sign and magnitude of the correlation between the data points, some quantitative information can be read off the correlation lines.
Assuming multivariate normal uncertainties, the expected values of the bins change when fixing one of the neighbouring bins according to \cite[P88]{Soch2025}
\begin{align}
    \expect[y_i | y_j] = \expect[y_i] + \sigma_i\rho_{ij} \frac{y_j - \expect[y_j]}{\sigma_j}.
\end{align}
This means that, given a single data point fluctuates to $+1\sigma$ of its uncertainty band, its neighbours' conditional distributions' expected value is shifted to the position where the correlation line is attached to the error bar.
In this case, the conditional variance of that data point is reduced from its total variance,
\begin{align}
    \variance[y_i|y_j] = \sigma_i^2 \qty(1 - \rho_{ij}^2).
\end{align}
This conditional variance does not depend on the actual value of the neighbouring bin $y_j$, it just requires that its value is fixed.

One can also rearrange this equation to yield the law of total variance for a data point \citep[P292]{Soch2025}:
\begin{align}
    \sigma_i^2 &= \expect[\variance[y_i|y_j]] + \variance[\expect[y_i|y_j]] \nonumber\\
        &= \variance[y_i|y_j] + \rho_{ij}^2 \sigma_i^2.\label{eq:totvar}
\end{align}
The left term of \autoref{eq:totvar}, the conditional variance, can be considered the \enquote{intrinsic} variance of the data point.
If the other data point is fixed to a specific value, this data point is still uncertain according to this variance.
The right part, proportional to the squared correlation coefficient, can be seen as the variance \enquote{caused} by the correlation with the other bin, and how its variation shifts the expected value around.

The position where the correlation lines attach to the error bars shows the portion of the uncertainty that is \enquote{explained} by the correlation with the neighbouring bin, while the total error bar shows the total.
This is similar to how sometimes error bars are shown split into a statistical and \enquote{statistical + systematic} part.
Of course, all of these relations are symmetric, and one can just as well interpret things as the variance of data point~$i$ explaining some some of the variance in data point~$j$, rather than the other way around.

This straight-forward interpretation of the correlation lines only works pairwise between neighbouring data points.
If we consider three neighbouring data points together and try to interpret how the outer points influence the middle, we lack the information of the correlations between the outer points to determine any quantitative things from the plot.
If the left bin is strongly correlated to the middle one, and the middle one is strongly correlated to the right one, then the left is likely to be correlated to the right one as well.
The exact strength of this is can only be determined if at least one of the displayed correlations is 100\%, though.
\enquote{Long-range} correlation effects that affect not just neighbouring bins require different ways of plotting them.

\section{Showing the Dominant Principal Component}

Depending on the type of data, the neighbouring correlations can be the most significant ones,
e.g., when the correlations are driven by detector resolution effects and their unfolding.
But in general, the correlation between neighbouring bins does not need to be the most significant.

A general way of identifying the most important components of a multivariate distribution is the Principal Component Analysis (PCA, see e.g., \cite{Abdi2010}).
The principal components are the $N$ unit-length eigenvectors $\vect{u}_i$ of the \emph{correlation} matrix, ordered by their corresponding eigenvalues $\lambda_i$ from largest to smallest.
If the uncertainties are uncorrelated, all eigenvalues will be 1.
In the presence of correlations, the eigenvalues will no longer be equal, but by construction they will always sum up to $N$.
The eigenvalues are a measure of how much each component contributes to the total variance in the uncertainty.

The correlation matrix $C$ of the covariance $V$ is just the sum of the contributions from these components:
\begin{align}
    C &= W^{-1} V W^{-1} \nonumber\\
      &= \sum_{i=1}^{N} \lambda_i \vect{u}_i \vect{u}_i^T,
\end{align}
where $W$ is a diagonal matrix of the square root of the diagonal elements of $V$.
To visualise how important the first component is to the total, one can reduce its contribution to get a remaining correlation matrix $K$:
\begin{align}
    K = C - s^2 \lambda_1 \vect{u}_1 \vect{u}_1^T.
\end{align}
Here, the scaling factor $s$ determines how much of the component is removed.

\begin{figure*}
    \centering
    \includegraphics[width=0.95\linewidth]{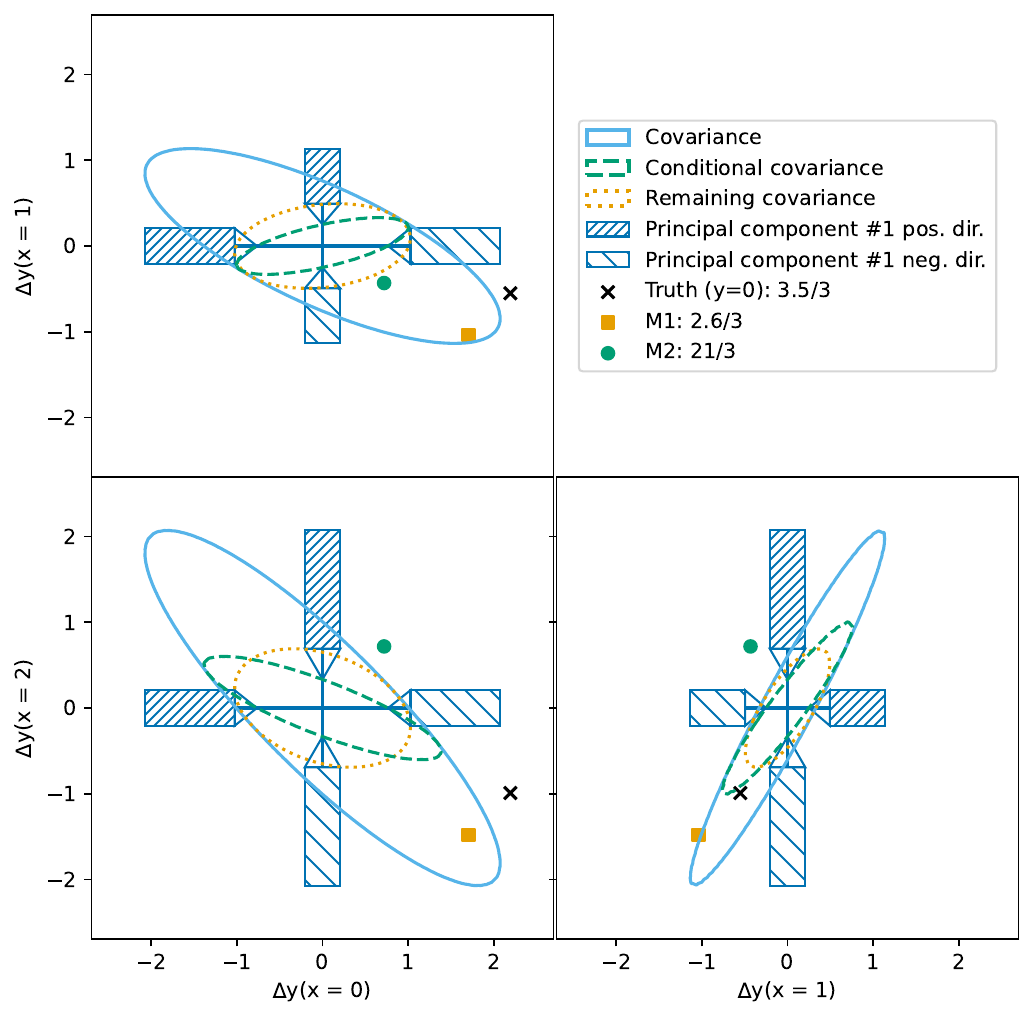}
    \caption{\label{fig:full} The example data shown in all pairwise 2D projections. The solid ellipses show the original covariance, i.e., where the M-distance of the marginalised uncertainties is 1. These ellipses determine the total length of the error bars, which correspond to the square root of the diagonal elements of the covariance matrix. The contribution of the first principal component of the covariance is reduced, and the remaining covariance is plotted as dotted lines. The diagonal elements of this remaining covariance determine the inner edges of the hatched areas. So the hatched areas show uncertainty that is caused by the first principal component. The type of hatching indicates whether the area is in the positive or negative direction of that component. The remaining covariance matrix still has correlations in it and for neighbouring bins those are indicated by the correlation lines in \autoref{fig:pcplot} (not shown here). The dashed ellipses show the conditional covariance of the full uncertainties when all other $\Delta y$ are assumed to be 0, i.e., it is a slice of the $N$-dimensional ellipsoid rather than a projection. Where these ellipses intersect the axes indicates the conditional uncertainty of that data point, assuming all others are fixed. This determines the position of the middle points of the triangles.}
\end{figure*}

When plotting the data with error bars corresponding to both $C$ and $K$, the difference between the error bars show the contribution that has been removed.
Importantly, the removed component stems from a single, 100\% correlated variation in the data along the direction of $W\vect{u}_1$.
So these differences have a defined direction in the $N$-dimensional data space.
This direction can be indicated by hatching the area between the full and remaining covariance errors differently depending on whether the error bars are on the same side as the corresponding component of $\vect{u}_i$ or the opposite.
This is done in \autoref{fig:pcplot}, and the concept is explained and sketched out in more detail in \autoref{fig:full}.

A full removal at $s = 1$ would mean that the remaining correlation matrix $K$ will be rank deficient,
and will thus be strongly correlated itself.
The aim is to make the plot less misleading and to make effects of correlations visible.
To that end, we need to aim to make the remaining covariance less correlated than the full covariance.
There are multiple choices for $s$ that can achieve this.

A conservative choice is to scale the contribution of the first principal component such that it is equal to the contribution of the second component.
To do this, one chooses $s^2 = (1 - \lambda_2 / \lambda_1)$.
The choice of the \enquote{target} eigenvalue is somewhat arbitrary, though,  and the ideal value depends on the plotted data.
A reasonable starting point is to aim for the median value: $s^2 = (1 - \lambda_m / \lambda_1)$.
If one wants to get the maximum effect in the plot, and assuming that the original covariance matrix is not rank deficient, one can also aim for the smallest eigenvalue: $s^2 = (1 - \lambda_N / \lambda_1)$.
Figures~\ref{fig:pcplot} and~\ref{fig:full} use the median target value, while Figures~\ref{fig:dpt} and~\ref{fig:ratio} use the smallest.

But no matter how $s$ is chosen, the remaining covariance will rarely be completely uncorrelated.
The remaining correlations can be partially included in the plot by plotting the remaining covariance with correlation lines as introduced in the previous section.

Highlighting the first principal component is most useful when this component is the dominant source of variance in the data, i.e., if the first eigenvalue of the correlation matrix is much larger than the others.
In this case, this component defines one particular direction in data space with much more freedom than any other directions.
A model should only be compared to this large uncertainty, if the difference to the data points aligns with this direction.
That means in the plots, the model prediction should be on the sides of the data points that share the same hatching pattern.
If the model does not align with the first principal component direction, it should be compared to the remaining covariance only.

Of course, it is also possible to show not just the first principal component, but the first $N_C$ components.
In this case, there will be $N_C$ scaling factors, which should be chosen so the target eigenvalue is the same for all of them.
$K$ is now the remaining covariance after subtracting the contributions of the $N_C$ components.
Hatched areas are then shown successively, by adding back the contributions from smallest to largest eigenvalue in turn, and filling the difference between the new and previous error bars.
Such plots can get very busy, though, and beyond the first or maybe second component, the readability drops sharply.
Examples of this are shown in appendix~\ref{sec:examples}.

\section{Conditional Uncertainties}

Another way to indicate the presence of correlations is to display the \emph{conditional} uncertainties for each data point as well as the marginal ones.
The conditional variance is the variance of a data point, assuming that all other data points are fixed.
If the uncertainties are multivariate normal -- which we often implicitly assume when they are parametrised as a covariance matrix -- the conditional variance is independent of the actual values of the other data points:
\begin{align}
    \sigma^2_{i,\text{cond}} = ((V^{-1})_{ii})^{-1}.
\end{align}
These conditional uncertainties of the full covariance $V$ are shown as the inner points of the triangles in Figures~\ref{fig:pcplot} and~\ref{fig:full}.

The conditional uncertainties represent slices through the parameter space (fixing other variables), as opposed to marginal uncertainties, which represent projections (ignoring other variables).
If the covariance matrix in question is uncorrelated, the conditional and marginal uncertainties are identical.
If the covariance is considerably more constrained in one directions compared to the others,
i.e., if one eigenvalue of the correlation matrix is much smaller than the others,
then the conditional uncertainties will correspond to the uncertainty in the direction of this strongest constraint.
The inclusion of the conditional uncertainties can thus give another indication of correlations and is complementary to the display of principal components.

\section{Combined Effect}

Compared to the current default style in \autoref{fig:diag}, plots in the style of \autoref{fig:pcplot} should enable authors to convey much more useful information about the shape of uncertainties in their data.
Naturally, this comes at the \enquote{cost} of making the plots more information dense and thus a bit harder to parse.
But the plots only contain additional information, and when in doubt, it is always possible to completely ignore the additional bits and only read off the outer limits of the hatched rectangles.
These correspond exactly to the marginal error bars shown in the default plot style.
For further interpretation of visual comparisons to model predictions, one can employ these rules of thumb:
\begin{itemize}
    \item Models should be compared to the outer edges of the hatched areas if the deviation from the data is on the sides with the same hatching style for \emph{all} data points, where it is outside the inner edges.
    \item Where the rectangles are too small to distinguish their fill style, the distinction does not matter.
    \item Otherwise, the models should be compared to the inner edges of the hatched areas.
    \item The inner triangle points show how much uncertainty of each data point is \enquote{intrinsic} and does not depend on the value of the other data points.
\end{itemize}
To help build an intuition how different correlation structures would look in this style of plot, examples are collected in Appendix~\ref{sec:examples}.

\section{A Real-World Example}
\label{sec:dpt}

\begin{figure*}
    \centering
    \includegraphics[width=\linewidth]{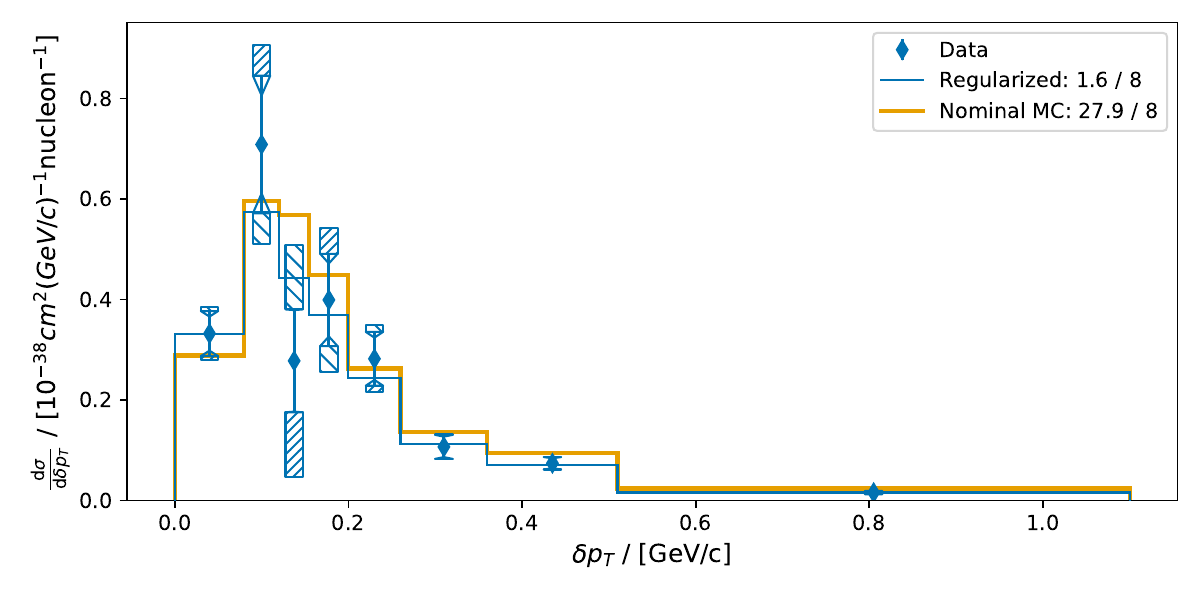}
    \caption{\label{fig:dpt}%
        The $\delta p_T$ result from \cite{Abe2018,Abe2018e} shown as a principal component plot.
        The nominal Monte Carlo model was provided by the original authors of the measurement.
        The strong anticorrelation of the second, third, and fourth bins are clearly visible in the first principal component.
        The regularised result is on the \enquote{right} side of all three data points, while the nominal model is not.
        The model thus \enquote{sees} only the remaining uncertainties without the first component.
        The correlation lines for the remaining covariance were omitted in this plot, since the remaining correlation was negligible and only reduced the readability of the data.%
    }
\end{figure*}

\autoref{fig:dpt} shows the result of the $\delta p_T$ cross-section measurement from \cite{Abe2018,Abe2018e} in the new principal component plot style.
The correlation matrix of this result is shown in Figures~\ref{fig:cor} and~\ref{fig:hinton}.
The hatched areas clearly show the considerable anticorrelations between the second, third, and fourth data points,
suggesting that the \enquote{dip} in that region is not physical, but a statistical fluctuation.

\begin{figure*}
    \centering
    \includegraphics[width=\linewidth]{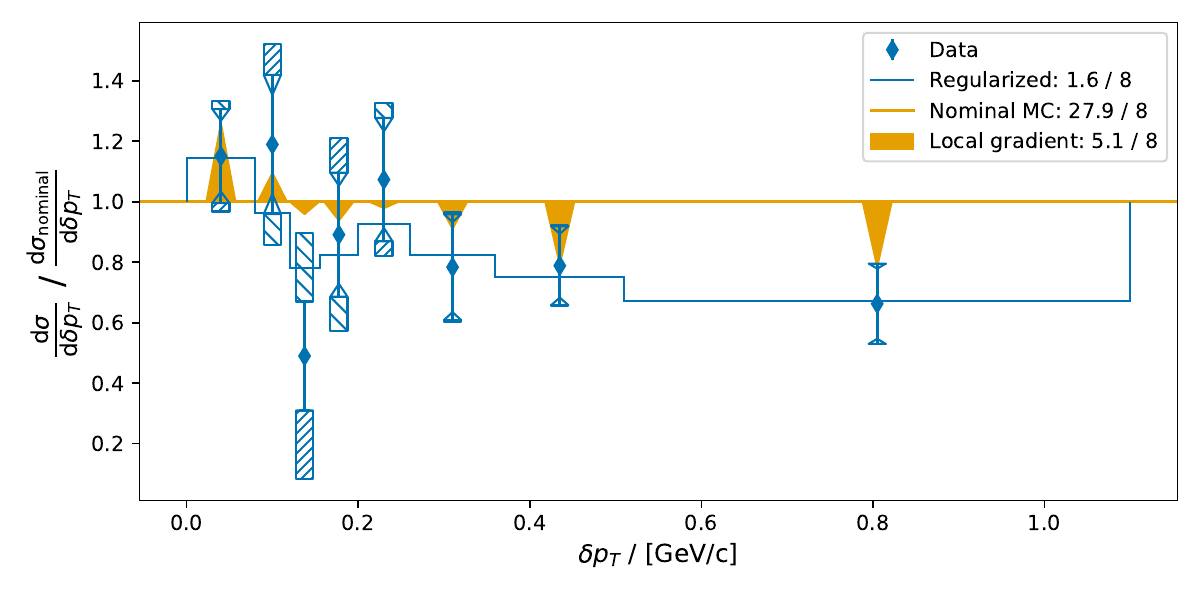}
    \caption{\label{fig:ratio}%
        The same data as in \autoref{fig:dpt} shown as a principal component plot combined with the model ratio and local gradient of the M-distance as proposed in \cite{Koch2022}.
        The gradient was scaled to minimise the squared M-distance at its endpoint, which is shown in the legend.
        Compared to \autoref{fig:dpt}, the anticorrelations of the first principal component are now better visible in the first 5 data points.
        Due to these correlations, the gradient is \emph{not} pointing directly at the data points.
        For one data point it is even pointing \emph{away} from the data, meaning the model fit to the data is improved by moving it away from that data point.%
    }
\end{figure*}

The error bars for the last data point in \autoref{fig:dpt} are too small to be seen,
which makes it impossible to judge how much that bin contributes to the bad agreement between the data and the nominal model.
This can be easily fixed by plotting the data to model ratio, as done in \autoref{fig:ratio}.
It also includes the local gradient of the Mahalanobis distance between the model and the data, as proposed in \cite{Koch2022}.
Together, this makes it clear that the discrepancy is actually more driven by the first and the two last bins,
even if the dip in the third bin might be visually the most significant deviation of the model from the data in \autoref{fig:dpt},

This is just a single example and cannot guarantee that this type of plot will always be useful in conveying information about the correlation between data points.
Every data set and every correlation structure is different, and an exhaustive analysis of when principal component plots make sense is beyond the scope of this paper.

\section{Conclusions}

The practice of plotting data points with only their marginal uncertainties can produce very misleading plots if there are considerable correlations in the covariance of the data.
The proposed plotting styles can add additional information to plots that make it easier to spot effects of such correlations.

The correlation lines are an easy way to include correlations between neighbouring data points.
In case there are no correlations, they will behave like a simple line connecting the data points.
The principal component plots can add information off the strength and direction of the largest principal components of the covariance.
The inclusion of conditional uncertainties in the plot can indicate the presence of strong constraints in the uncertainty that are not reflected in the size of the marginal error bars.
In case there are no correlations, they will revert to look like the usual \enquote{marginal error only} plots.
In either case, the correlation information is purely additive, and can easily be ignored, yielding the same information that the classical plots provide.

These new plotting styles are still not able to show the full information of all correlations in the data in the general case.
The only practical way to show all correlation information is to plot the correlation matrix directly.
Hinton plots are a good way to make these plots readable also in the absence of colour information, e.g., for colour-blind-friendly publications, or black-and-white prints.
This has a huge impact on the accessibility of research,
as hundreds of millions of people world wide suffer from some form of colour vision deficiency.

As a rule of thumb, it is usually a good idea to provide a plot of the full correlation matrix in the form of a Hinton diagram.
It contains the full correlation information between all dimensions of the data, even if it can be a bit hard to parse.
Then, depending on the actual structure of the data and correlations, the correlation lines or principal component plots can be used to insert additional information into the data plot itself.
The choice of which to use depends on the data, and it is difficult to make general judgements.
One should try and check which plotting style adds the most information without overloading the plot.
Again, as a rule of thumb, if there is a single dominating principal component, the PC plot will probably give good results.
If the data points are not too \enquote{squished together} in the $x$-direction and correlations are mostly short range, the correlation lines will likely add useful information.

\section*{Code and data availability}

The plotting styles proposed in this paper are implemented in the \pkg{NuStatTools} \proglang{Python} package, and available online \citep{Koch2026}. The plots in this document were generated with version 0.8.0 of the package. The \code{Jupyter Notebook} with the actual plotting commands for the artificial examples is available as supplementary material. The data shown in \autoref{sec:dpt} is available in their data release \citep{Abe2018e}.

\section*{Acknowledgements}

I would like to thank London Cooper-Troendle for the stimulating discussion at NuInt 2025 that kick-started this work, as well as for further discussions and feedback later on.
This work was funded by the Deutsche Forschungsgemeinschaft (DFG, German Research Foundation) under Germany’s Excellence Strategy – EXC 2118 PRISMA+ – 390831469.

\bibliography{biblio}

\clearpage
\appendix

\section{Illustrative Examples}
\label{sec:examples}

All figures in this appendix use the median eigenvalue as target value for the principal component reduction.

\begin{figure*}
    \centering
    \subfloat[No correlation]{\includegraphics[width=\pw]{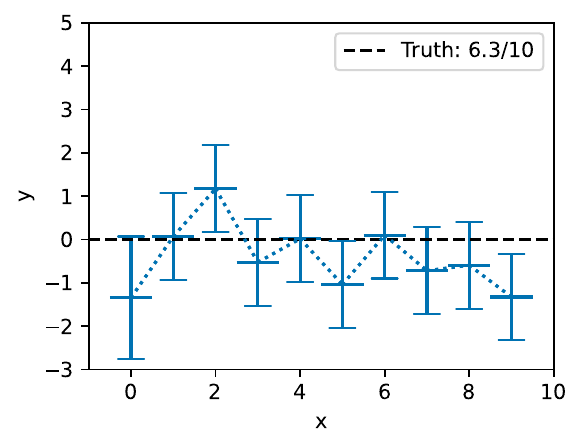}\label{fig:ex-uncor}}
    \subfloat[Sum constrained]{\includegraphics[width=\pw]{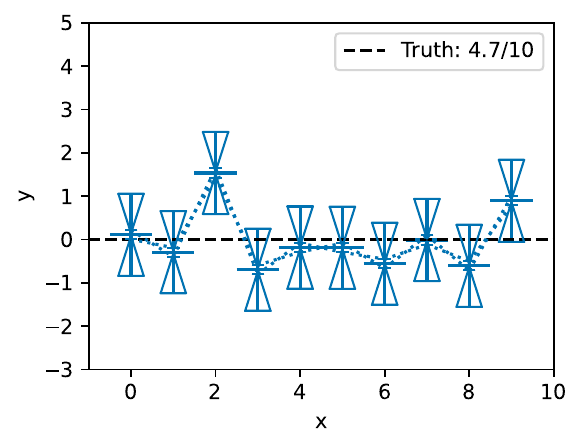}\label{fig:ex-shape}}\\[1em]
    \subfloat[Single dominating uncertainty]{\includegraphics[width=\pw]{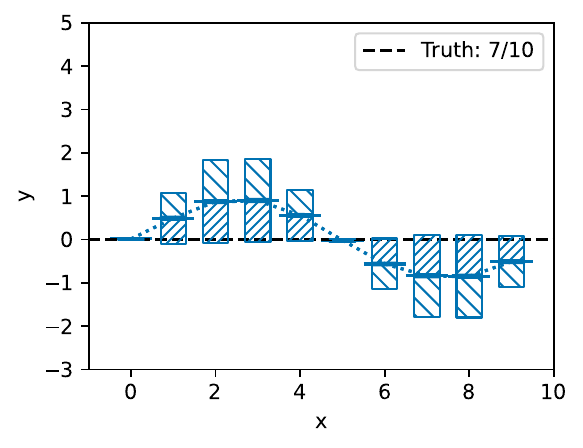}\label{fig:ex-single}}
    \subfloat[Two disjunct uncertainties]{\includegraphics[width=\pw]{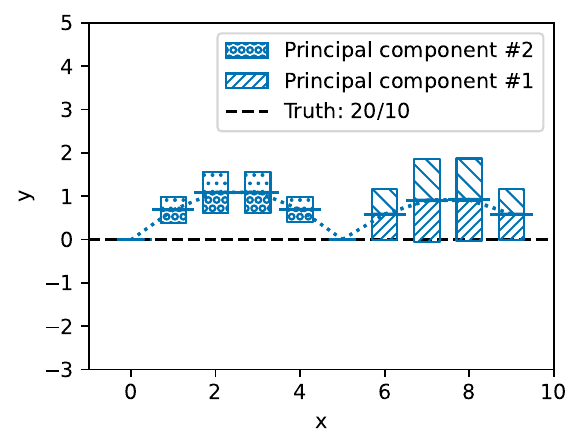}\label{fig:ex-double}}\\[1em]
    \subfloat[Three overlapping uncertainties]{\includegraphics[width=\pw]{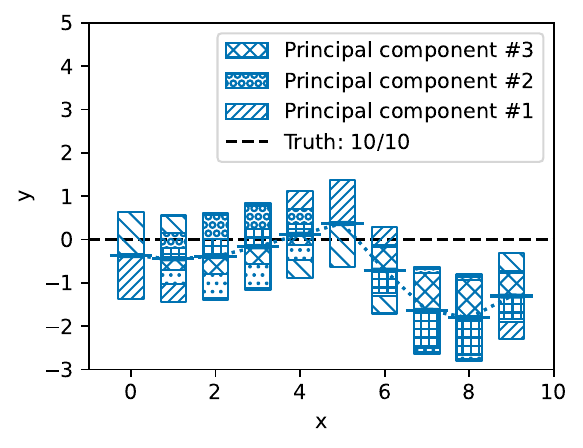}\label{fig:ex-triple}}
    \subfloat[First component of three overlapping uncertainties]{\includegraphics[width=\pw]{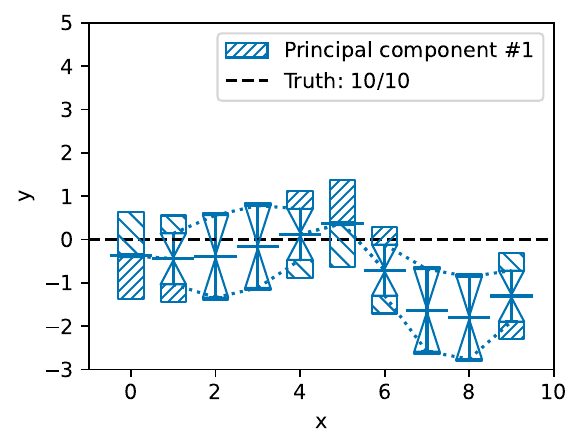}\label{fig:ex-triple-one}}
    \caption{Illustrative example plots}
\end{figure*}

\autoref{fig:ex-uncor} shows that in the case of perfectly uncorrelated data, the proposed principal component plot looks like a regular plot of only the diagonal elements of the covariance.
The reason for this is that for uncorrelated data, the correlation matrix will be the identity matrix and all eigenvalues are 1.
Since all eigenvalues are identical, nothing gets removed and $K$ is identical to the full covariance.

\autoref{fig:ex-shape} shows data where there is only very weak negative correlation between neighbouring bins.
There is no visible principal component and the correlation lines are not very far off from connecting the centre points.
But the covariance is constructed such that it only has shape freedom.
The sum of the data points is perfectly constrained.
This is visible in the conditional uncertainties, which are very small.

\autoref{fig:ex-single} shows a principal component plot for data with uncertainty that is completely dominated by a single principal component.
Basically the whole error bars are made up from this one component,
showing that there is no freedom to move any data point without moving all the others.

\autoref{fig:ex-double} shows a principal component plot for data with uncertainty that is completely dominated by two principal components, that act on different data points.
Each \enquote{lobe} is perfectly correlated and has no shape freedom, but the two lobes are independent of one another.

\autoref{fig:ex-triple} shows a principal component plot for data with uncertainty that is completely dominated by three overlapping principal components.
At this point, it is very difficult to glean any useful information from plotting all three components.
In cases like this, it might be better to plot only the first principal component and indicate that there is more going on with the correlation lines and conditional uncertainties, like is shown in \autoref{fig:ex-triple-one}.

\clearpage

\end{document}